\documentclass[aps,prl,twocolumn,groupedaddress,floatfix]{revtex4}

\usepackage{graphics}
\usepackage{color}
\usepackage{siunitx}

\begin{document}

\title{Testing {\it{ab initio}} nuclear structure in neutron-rich nuclei:  \\
lifetime measurements of second 2$^+$ states in $^{16}$C and $^{20}$O}



\author{M. Ciema\l a$^{1}$\footnote{Corresponding authors: Michal.Ciemala@ifj.edu.pl,\\Bogdan.Fornal@ifj.edu.pl, Silvia.Leoni@mi.infn.it}, 
S. Ziliani$^{2,3}$, F.C.L. Crespi$^{2,3}$, S. Leoni$^{2,3,*}$, B. Fornal$^{1,*}$, A. Maj$^{1}$,  P. Bednarczyk$^{1}$, G. Benzoni$^{3}$,\\
A. Bracco$^{2,3}$, C. Boiano$^{3}$, S. Bottoni$^{2,3}$, S. Brambilla$^{3}$, M. Bast$^{4}$, M.  Beckers$^{4}$, T. Braunroth$^{4}$, F. Camera$^{2,3}$,\\ 
N. Cieplicka-Ory\'nczak$^{1}$, E. Cl\'ement$^{5}$, S. Coelli$^{3}$, O. Dorvaux$^{6}$, S. Erturk$^{7}$, G. de France$^{5}$, C. Fransen$^{4}$, A.  Goldkuhle$^{4}$,\\
J. Grebosz$^{1}$, M.N. Harakeh$^{8}$, \L.W. Iskra$^{1,3}$, B. Jacquot$^{5}$, A. Karpov$^{9}$, M. Kici\'nska-Habior$^{10}$, Y. Kim$^{5}$, M. Kmiecik$^{1}$,\\
 A. Lemasson$^{5}$, S.M. Lenzi$^{11,12}$, M. Lewitowicz$^{5}$, H. Li$^{5}$,
 I. Matea$^{13}$, K. Mazurek$^{1}$, C. Michelagnoli$^{14}$,\\
 M. Matejska-Minda$^{15,1}$, B. Million$^{3}$, C. M\"uller-Gatermann$^{4}$,  V. Nanal$^{16}$, P. Napiorkowski$^{15}$, D.R. Napoli$^{17}$,
  R. Palit$^{16}$,\\
   M. Rejmund$^{5}$, Ch. Schmitt$^{6}$, M. Stanoiu$^{18}$, I. Stefan$^{13}$, E. Vardaci$^{19}$ , B. Wasilewska$^{1}$, O. Wieland$^{3}$, M. Zieblinski$^{1}$,\\
    M. Zieli\'nska$^{20}$, A. Ata\c{c}$^{21}$, D. Barrientos$^{22}$, B. Birkenbach$^{4}$, A.J. Boston$^{23}$, B. Cederwall$^{21}$, L. Charles$^{6}$, J. Collado$^{24}$,\\
D.M. Cullen$^{25}$, P. D\'esesquelles$^{26}$, C. Domingo-Pardo$^{27}$, J. Dudouet$^{28}$, J. Eberth$^{4}$, V. Gonz\'alez$^{24}$, J. Goupil$^{5}$,\\
 L.J. Harkness-Brennan$^{23}$, H. Hess$^{4}$, D.S. Judson$^{23}$, A. Jungclaus$^{29}$, W. Korten$^{20}$, M. Labiche$^{30}$, A. Lefevre$^{5}$,\\ R. Menegazzo$^{11}$,
D. Mengoni$^{11,12}$, J. Nyberg$^{31}$, R.M. Perez-Vidal$^{27}$, Zs. Podolyak$^{32}$, A. Pullia$^{2,3}$, F. Recchia$^{11,12}$,\\ P. Reiter$^{4}$, F. Saillant$^{5}$, M.D. Salsac$^{20}$, E. Sanchis$^{24}$, O. Stezowski$^{28}$, Ch. Theisen$^{20}$, J.J. Valiente-Dob\'on$^{17}$,\\
J.D. Holt$^{33}$, J. Men\'endez$^{34,35}$, A. Schwenk$^{36,37,38}$, J. Simonis$^{39}$
}  

%\email[]{Silvia.Leoni@mi.infn.it}
%\thanks{}

\affiliation{$^1$ Institute of Nuclear Physics, PAN, 31-342 Krak\'ow, Poland}

\affiliation{$^2$ Dipartimento di Fisica, Universit$\grave{a}$ degli Studi di Milano, I-20133 Milano, Italy}

\affiliation{$^3$ INFN sezione di Milano, via Celoria 16, 20133, Milano, Italy}

\affiliation{$^4$ Institut f\"ur Kernphysik, Universit\"at zu K\"oln, 50937 Cologne, Germany}

\affiliation{$^5$ GANIL, CEA/DRF-CNRS/IN2P3, Bd. Henri Becquerel, BP 55027, F-14076 Caen, France}

\affiliation{$^6$ CNRS/IN2P3, IPHC UMR 7178, F-67037 Strasbourg, France}

\affiliation{$^7$ Nigde Omer Halisdemir University, Science and Art Faculty, Department of Physics, Nigde, Turkey}

\affiliation{$^8$ KVI - Center for Advanced Radiation Technology, Groningen, Netherlands}

\affiliation{$^{9}$ FLNR, JINR, 141980 Dubna, Russia}

\affiliation{$^{10}$ Faculty of Physics, University of Warsaw, Warsaw, Poland }

\affiliation{$^{11}$ INFN Sezione di Padova, I-35131 Padova, Italy}

\affiliation{$^{12}$ Dipartimento di Fisica e Astronomia, Universit$\grave{a}$ degli Studi di Padova, I-35131 Padova, Italy}

\affiliation{$^{13}$ IPN Orsay Laboratory, Orsay, France}

\affiliation{$^{14}$ Institut Laue-Langevin (ILL), Grenoble, France}

\affiliation{$^{15}$ Heavy Ion Laboratory, University of Warsaw, PL 02-093 Warsaw, Poland}

\affiliation{$^{16}$ Tata Institute of Fundamental Research, Mumbai 400005, India}

\affiliation{$^{17}$ INFN Laboratori Nazionali di Legnaro, I-35020 Legnaro, Italy}

\affiliation{$^{18}$ IFIN-HH, Bucharest, Romania}

\affiliation{$^{19}$ Universit$\grave{a}$ degli Studi di Napoli and INFN sez. Napoli, Italy}

\affiliation{$^{20}$ IRFU, CEA/DRF, Centre CEA de Saclay, F-91191 Gif-sur-Yvette Cedex, France}

\affiliation{$^{21}$ Department of Physics, Royal Institute of Technology KTH, SE-10691 Stockholm, Sweden}

\affiliation{$^{22}$ CERN, CH-1211 Geneva 23, Switzerland}

\affiliation{$^{23}$ Oliver Lodge Laboratory, The University of Liverpool, Liverpool, L69 7ZE, UK}

\affiliation{$^{24}$ Departamento de Ingenier\'ia Electr\'onica, Universitat de Valencia, Burjassot, Valencia, Spain}

\affiliation{$^{25}$ Nuclear Physics Group, Schuster Laboratory, University of Manchester, Manchester, M13 9PL, UK}

\affiliation{$^{26}$ CSNSM, Universit\'e Paris-Sud/CNRS-IN2P3, Universit\'e Paris-Saclay, Bat 104, F-91405 Orsay Campus, France}

\affiliation{$^{27}$ Instituto de F\'isica Corpuscular,  CSIC-Universidad de Valencia, E-46071 Valencia, Spain}

\affiliation{$^{28}$ Universit\'e de Lyon 1, CNRS/IN2P3, UMR5822, IPNL, F-69622 Villeurbanne Cedex, France}

\affiliation{$^{29}$ Instituto de Estructura de la Materia, CSIC, Madrid, Spain}

\affiliation{$^{30}$ STFC Daresbury Laboratory, Daresbury, Warrington, WA4 4AD, UK}

\affiliation{$^{31}$ Department of Physics and Astronomy, Uppsala University, SE-75120 Uppsala, Sweden}

\affiliation{$^{32}$ Department of Physics, University of Surrey, Guildford, GU2 7XH, UK}

\affiliation{$^{33}$ TRIUMF, 4004 Wesbrook Mall, Vancouver BC V6T 2A3, Canada}

\affiliation{$^{34}$ Center for Nuclear Study, University of Tokyo, Tokyo 113-003, Japan}

\affiliation{$^{35}$ Departament de F\'isica Qu\'antica i Astrof\'isica Mart\'i i Franqu\'es 1, 08028 Barcelona, Spain}

\affiliation{$^{36}$ Institut f\"ur Kernphysik, Technische Universit\"at Darmstadt, 64289 Darmstadt, Germany}

\affiliation{$^{37}$ ExtreMe Matter Institute EMMI, GSI Helmholtzzentrum f\"ur Schwerionenforschung GmbH, 64291 Darmstadt, Germany}

\affiliation{$^{38}$ Max-Planck-Institut f\"ur Kernphysik, Saupfercheckweg 1, 69117 Heidelberg, Germany}

\affiliation{$^{39}$ Institut f\"ur Kernphysik and PRISMA Cluster of Excellence, Johannes Gutenberg-Universit\"at Mainz, 55128 Mainz, Germany}

\date{\today}

\begin{abstract}
{To test the predictive power of {\it{ab initio}} nuclear structure theory, the lifetime of the second 2$^{+}$ state in neutron-rich $^{20}$O, $\tau$(2$^+_2$) = 150$^{+ 80}_{-30}$ fs, and an estimate for the lifetime of the second 2$^+$ state in $^{16}$C have been obtained, for the first time. The results were achieved via a novel Monte Carlo technique that allowed us to measure nuclear state lifetimes in the tens-to-hundreds femtoseconds range, by analyzing the Doppler-shifted $\gamma$-transition line shapes of products of low-energy transfer and deep-inelastic processes in the reaction $^{18}$O (7.0 MeV/u) + $^{181}$Ta. The requested sensitivity could only be reached owing to the excellent performances of the AGATA $\gamma$-tracking array, coupled to the PARIS scintillator array and to the VAMOS++ magnetic spectrometer. The experimental lifetimes agree with predictions of \textit{ab initio} calculations using two- and three-nucleon interactions, obtained with the valence-space in-medium similarity renormalization group for $^{20}$O, and with the no-core shell model for $^{16}$C. The present measurement shows the power of electromagnetic observables, determined with high-precision $\gamma$ spectroscopy, to assess the quality of first-principles nuclear structure calculations, complementing common benchmarks based on nuclear energies. The proposed experimental approach will be essential for short lifetimes measurements in unexplored regions of the nuclear chart, including r-process nuclei, when intense ISOL-type beams become available.}
\end{abstract}

%\pacs{XXX}

%\keywords{}
\maketitle
%\section{\label{intro}}

Atomic nuclei are composed of protons and neutrons, which, in turn, are systems of quarks and gluons confined via the strong interaction, as described by Quantum Chromodynamics (QCD). Ideally, one would like to obtain the properties of nuclei solving QCD, but despite recent progress this is beyond current computational capabilities \cite{Bea11, Aok12, Det19}. At the energy and momentum scales relevant for nuclear structure, chiral effective field theory (EFT), an effective theory based on the symmetries of QCD, provides a practical alternative \cite{Epe09, Mac11, Ham13}. In chiral EFT the degrees of freedom are nucleons and pions, and nuclear forces are given by a systematic expansion of two- (NN), three- (3N) and many-nucleon interactions that includes pion exchanges and contact terms. 

In recent years, chiral EFT interactions have been combined with \textit{ab initio} approaches that consider all nucleons in the solution of the nuclear many-body problem, in studies of mid-mass nuclei up to tin \cite{Hag14, Heb15, Nav16, Lee16, Her17, Str19}. Theoretical results are typically compared to the simplest experimental observables, namely binding and excitation energies. First calculations of medium-mass nuclei with chiral NN+3N interactions predicted correctly the oxygen neutron dripline at $^{24}$O \cite{Heb15,Ots10}, and later works reproduced well the excitation spectra of neutron-rich oxygen isotopes \cite{Heb15}. In neutron-rich calcium and nickel isotopes, the shell evolution was also satisfactorily predicted \cite{Wie13, Ste13, Tan19}. Tests of \textit{ab initio} calculations against other observables include charge radii \cite{Gar16, Lap16}, beta decay lifetimes \cite{Gys19}, and elastic proton scattering off $^{10}$C \cite{Kum17}. Electromagnetic (EM) responses have also been studied in selected nuclei \cite{Bac14, Rai19, Sim19}. A general agreement between theory and experiment was found. 

Electric and magnetic $\gamma$ decays provide a more demanding, complementary test of theoretical approaches. So far, EM decays in light/medium-mass systems have only been explored in few cases \cite{McC09,McC12,Par17, Gar17, Hen18, Ruo19}. In contrast to energies or beta decays, \textit{ab initio} methods do not yet consistently describe
all the data related to EM transitions. This calls for precision measurements of EM observables which are sensitive to the details of the calculations. Ideal cases are neutron-rich O and C isotopes. Here, \textit{ab initio} approaches show a strong sensitivity of EM decays to 3N forces which significantly affect the lifetime of selected excited states, by changing the wave function composition \cite{For13}.

In this paper, we focus on the lifetime determination of the second 2$^+$ excitations, 2$^+_2$, in $^{20}$O and $^{16}$C nuclei. We confront our results with predictions of the valence-space in-medium similarity renormalization group (VS-IMSRG) and of the no-core shell model (NCSM) of Ref. \cite{For13}, both including NN and 3N forces. Previous experiments have only set limits on the lifetime of these 2$^+_2$ states, of the order of a few ps, and provided information on branching ratios in their decays \cite{Wie05, Wie08, Pet12, Vos12}. In this work, we aim at a much more stringent test of \textit{ab initio} calculations by measuring the 2$^+_2$ state lifetimes. 

In $^{20}$O, the first-excited 2$^+_1$ state at 1674 keV decays
with a lifetime of 10.4(9) ps \cite{Pri16}. The 2$^+_2$ state, of our interest, located at 4070 keV, was found to decay to the 2$^+_1$ state with 72(8)$\%$ branch \cite{Wie05}, in parallel to the direct ground state branch for which contradicting B(E2) information is reported from intermediate-energy Coulomb excitation \cite{Nak17,Try03}. In $^{16}$C, the 2$^+_1$ state at 1762 keV decays with a lifetime of 11.4(10) ps \cite{Pet12}, while for the 2$^+_2$ state lifetime only the upper limit of 4 ps is known \cite{Wie08, Pet12}. 
Theoretical predictions suggest that the lifetimes of these 2$^+_2$ states are in the tens-to-hundreds femtoseconds range. This poses an experimental challenge since such neutron-rich systems can only be produced, with sizable cross sections, in: \textit{i)} relativistic heavy-ion fragmentation, for which the lower limit in lifetime determination is few hundreds femtoseconds, as shown by Morse et al. \cite{Mor18a}, \textit{ii)} low-energy transfer and deep-inelastic reactions, where the complex structure of the product velocity distribution, caused by large energy dissipations \cite{Zag14, Kar17}, does not allow to use standard Doppler Shift Attenuation Methods \cite{Nol79}.

\begin{figure}[ht]
%\vspace*{0.1cm}
\resizebox{0.5\textwidth}{!}{\includegraphics{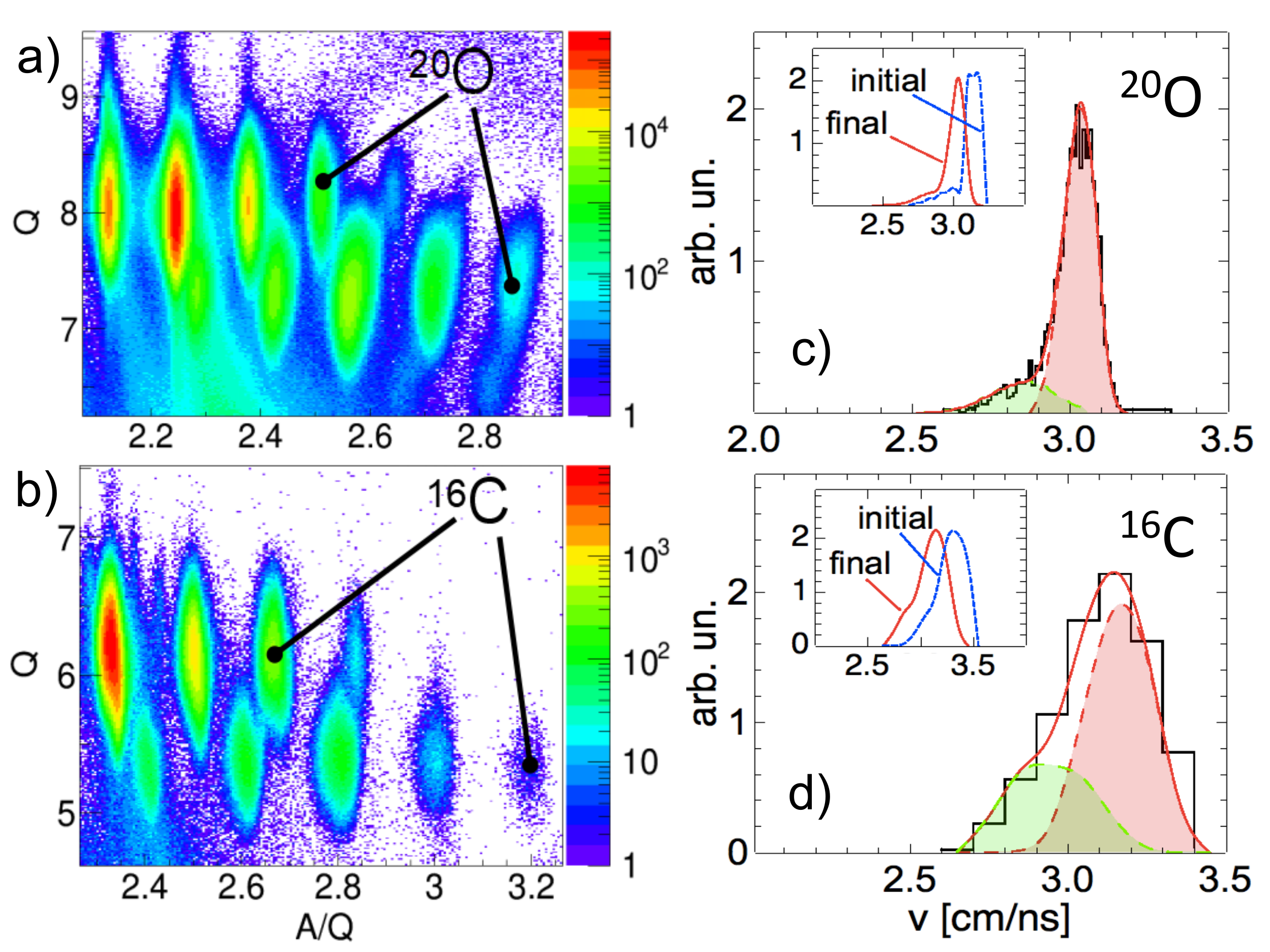}}
\caption{\textit{Left}: Identification plots, charge Q versus A/Q, for O (a) and C (b) ions, as measured by VAMOS++. \textit{Right}: velocity distributions measured at the VAMOS++ focal plane (black histograms), in coincidence with 
$2^+_2\rightarrow2^+_1$ transitions of $^{20}$O (c) and $^{16}$C (d), detected in AGATA. Solid red lines are simulated final velocity distributions - quasi-elastic (deep-inelastic) components are given by red (green) shaded areas. Insets: reconstructed initial (dashed blue) and final (solid red lines) product velocities.} 
\label{Fig1}
\end{figure}

In this work, we have developed a technique which enables us to access tens-to-hundreds femtoseconds lifetimes in exotic neutron-rich nuclei, with low-energy transfer and deep-inelastic heavy-ion reactions – our approach is a significant extension of the Doppler-Shift Attenuation Method to cases without well-defined reaction kinematics. It will be an essential and quite unique tool to determine short lifetimes in nuclei with large neutron excess, including r-process nuclei, when intense ISOL-type beams \cite{ISOL}, currently under development, are available. The novelty of the method relies on the accurate reconstruction of the complex initial velocity distribution of the reaction product excited to a specific nucler state, including contributions from direct and dissipative processes. The Doppler-shifted $\gamma$-transition line shape is then simulated considering the precisely determined detection angle between the $\gamma$-ray and the reconstructed initial product velocity inside the target. It will be shown that the required sensitivity to the lifetimes could only be achieved by the excellent performances of our integral AGATA+PARIS+VAMOS detection system.

In the present study, $^{16}$C and $^{20}$O nuclei were populated in both direct transfer and deep-inelastic processes induced by an $^{18}$O beam at 126 MeV on a $^{181}$Ta target (6.64 mg/cm$^2$). The beam energy at the center of the target was $\sim$116 MeV, i.e., $\sim$50$\%$ above the Coulomb barrier, and projectile-like products had velocity v/c$\sim$10$\%$. The experiment was performed at GANIL with 31 High-Purity Ge detectors of the AGATA array \cite{Akk12,Cle17}, coupled to a scintillator array consisting of two large volume 
(3.5"$\times$8") LaBr$_3$ detectors plus two clusters of the PARIS setup \cite{Maj09}, which produced excellent time reference for the reaction.  Reaction products were detected in the VAMOS++ spectrometer \cite{Rej11,Pul08}, placed at the reaction grazing angle of \ang{45} (opening angle $\pm$\ang{6}) with respect to the beam direction and aligned with the center of AGATA. Relative to the VAMOS++ axis, AGATA covered the {\ang{115}-\ang{175}} angular range, while the scintillators were placed at \ang{90}. More than 10$^7$ events were collected requiring coincidences between projectile-like products detected in VAMOS++ and $\gamma$ rays in AGATA or PARIS. The VAMOS++ setting was optimized to detect $^{20}$O within a large velocity range, including quasi-elastic and deep-inelastic processes. Other products with charge 
5$\leq$ Z $\leq$9 and mass number 11$\leq$ A $\leq$21 were also detected. Identification plots of O and C ions are given in Fig. \ref{Fig1}, together with the velocity distributions measured in VAMOS++ for the 2$^+_2$ states in $^{20}$O and $^{16}$C. In the case of 
$^{20}$O, the velocity distribution shows a pronounced peak, corresponding to the direct population in a quasi-elastic process, while the tail, extending towards lower velocities, is associated with higher kinetic energy loss. The velocity distribution is similar in $^{16}$C, although with a much larger contribution ($\sim32\%$) from dissipative processes.

\begin{figure}[ht]
%\vspace*{0.1cm}
\resizebox{0.48\textwidth}{!}{\includegraphics{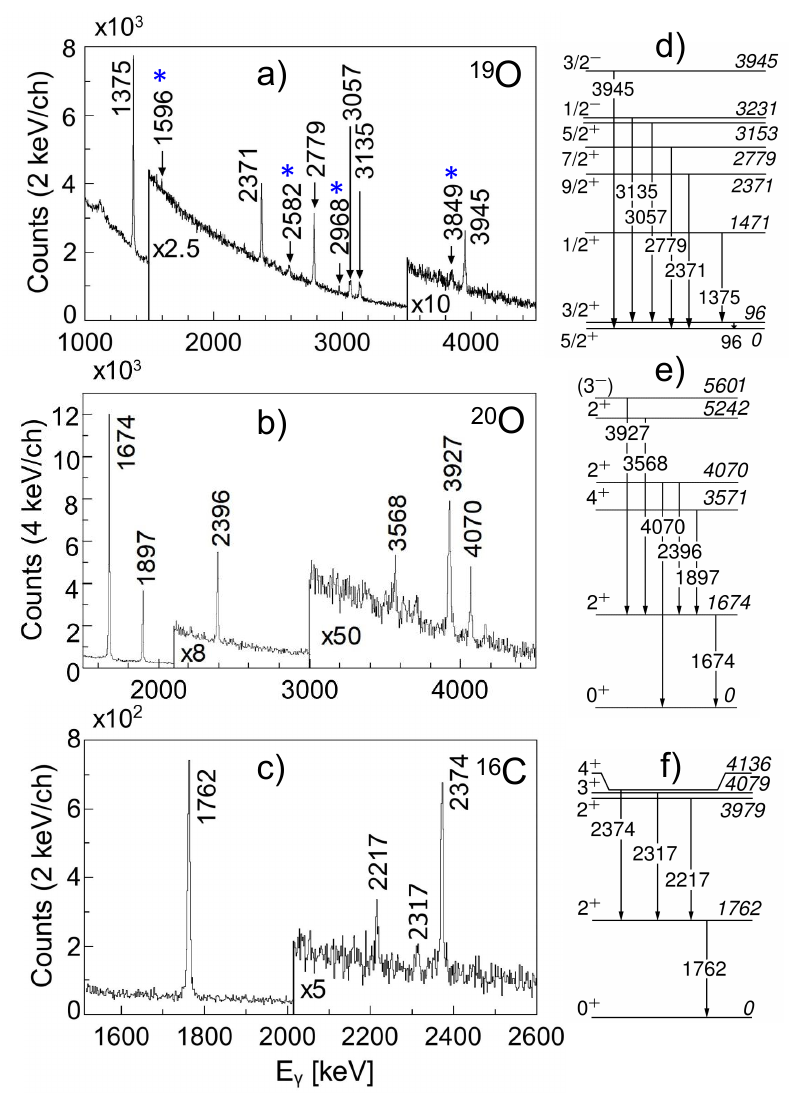}}
\caption{\textit{Left}: $\gamma$-ray spectra of $^{19}$O (a), $^{20}$O (b) and $^{16}$C (c), as measured by AGATA. \textit{Right}: Corresponding level schemes of $^{19}$O (d), $^{20}$O (e) and $^{16}$C (f). In a), blue stars mark weak transitions of $^{19}$O not shown in d). } 
\label{Fig2}
\end{figure}

Portions of $\gamma$-ray spectra obtained with AGATA by gating on $^{19}$O, $^{20}$O and $^{16}$C ions, Doppler-corrected considering the product velocity vector measured in VAMOS++, are shown in Fig. \ref{Fig2}, panels a), b) and c), respectively. All visible $\gamma$ rays correspond to known transitions (see level schemes on the right). A closer inspection reveals that some of the peaks are narrow and their energies agree, within uncertainty, with earlier studies. This is the case of transitions from relatively long-lived levels ($\tau>$1 ps), emitted in flight outside the target, as for example the 1375-keV and 2371-keV lines in $^{19}$O, the 1674-keV line in $^{20}$O and the 1762-keV peak in $^{16}$C \cite{Nudat}.

\begin{figure*}[ht]
%\vspace*{0.1cm}
\resizebox{1.0\textwidth}{!}{\includegraphics{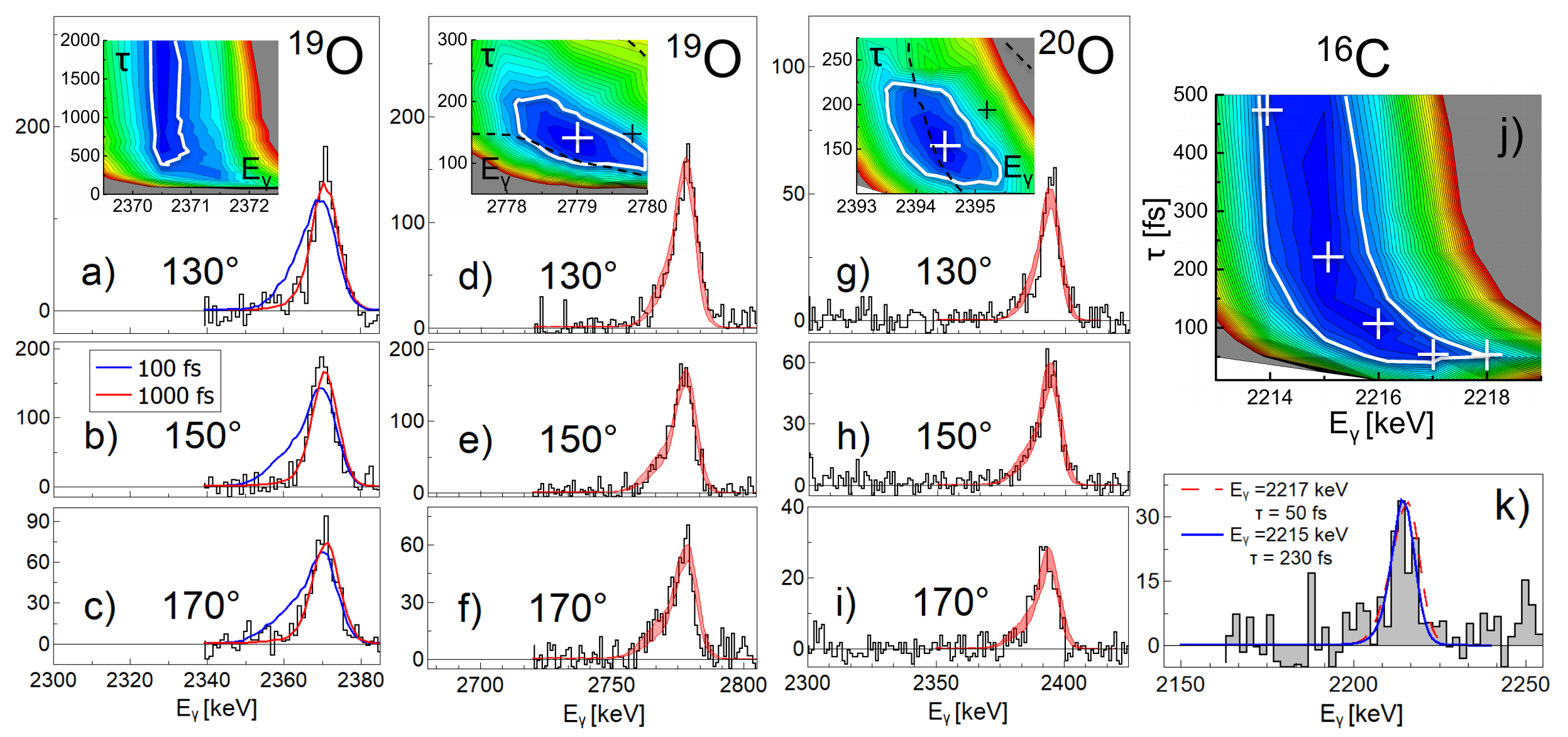}}
\caption{\textit{Panels (a)-(c), (d)-(f) and (g)-(i)}: 2371-, 2779- and 2396-keV $\gamma$ rays from the long-lived 
($\tau>$3.5 ps) 9/2$^+$ state in $^{19}$O, and the 7/2$^+$ and the 2$^+_2$ states in $^{19}$O and $^{20}$O, respectively, as measured by AGATA (precision $<$4 mm) in the angular ranges \ang{130}$\pm$\ang{10}, \ang{150}$\pm$\ang{10} and \ang{170}$\pm$\ang{10}. In (d)-(f) and (g)-(i) simulated spectra (shaded bands) are calculated on basis of the global $\chi^2$ lifetime-energy surface (see corresponding insets), with 1$\sigma$ uncertainty (white contour line) around the optimum $\gamma$-energy ($E_{\gamma}$) and lifetime ($\tau$) values (white cross). Black crosses and dashed lines indicate the $\chi^2$ minimum and 1$\sigma$ uncertainty for $\gamma$-detection angles defined by the AGATA front segment centers (precision $\sim$20 mm). In (a)-(c), simulated spectra for $\tau$ = 100 and 1000 fs (blue and red lines, respectively) are shown to demonstrate the absence of broadening and tails for decays from long-lived states. \textit{Panel j)}: $\chi^2$ surface for the $^{16}$C 2$^+_2\rightarrow2^+_1$ transition. \textit{Panel k)}: $^{16}$C  2$^+_2\rightarrow2^+_1$  transition measured over the full AGATA angular range (histogram), compared with simulated spectra relative to the minimum of the $\chi^2$ map, for 
$E_{\gamma}$ = 2215 keV ($\tau$=230 fs) and $E_{\gamma}$ = 2217 keV ($\tau$=50 fs).} 
\label{Fig3}
\end{figure*}

In contrast, other lines, depopulating states with lifetimes shorter than 1 ps, exhibit rather large widths and tails. This indicates that the corresponding $\gamma$ rays were partly emitted during the stopping process of the reaction product inside the target, i.e., at larger velocity than the one measured in VAMOS++. In these cases, the Doppler-broadened line shape can be used to determine the lifetime of the state. 

The Monte Carlo simulation of the Doppler-shifted $\gamma$-transition line shapes was performed in
three steps \cite{Cie20}. First, we reconstruct, iteratively, the initial product velocity distribution inside the target, associated with the population of a given state (see Fig. \ref{Fig1}). The procedure is based on the VAMOS++ measured velocity-angle
distribution (the spectrometer response function is considered \cite{Pul08}), the reaction kinematics calculated for the selected state (including direct population and more dissipative processes), a random 
probability of reaction occurrence over the target thickness, and the slowing-down and straggling processes inside the target, from Ziegler et al. \cite{Zie85,Ann88}. Second, we simulate the Doppler-shifted energy measured in AGATA for a transition emitted by the moving product
nucleus, with lifetime and transition energy as parameters. Here, the angle between the product velocity at the emission point and the 
$\gamma$-ray direction is determined with precision of $\sim$\ang{1.5} (i.e., \ang{1} for AGATA and \ang{1} for VAMOS++). Finally, 
we minimize the $\chi^2$ surface, in lifetime-energy coordinates, by comparing the simulated and experimental transition line shapes, for three selected angular ranges (i.e., {\ang{120}-\ang{140}}, {\ang{140}-\ang{160}} and {\ang{160}-\ang{180}}), simultaneously.

Figures \ref{Fig3}(a)-(c) and (d)-(f) show the 2371- and 2779-keV lines from the decay of the 9/2$^+$ and 7/2$^+$ states in 
$^{19}$O, with lifetimes $\tau>$3.5 ps and  $\tau$ = 92(19) fs, respectively \cite{Bro71,Hib71}. In Fig. \ref{Fig3}(d)-(f), the spectra for each angular range are overlaid with the results of the simulations, within 1$\sigma$ uncertainty around the optimum $\gamma$ energy ($E_{\gamma}$) and lifetime value. The insets show the corresponding $\chi^2$ surface, with white cross and solid line indicating the well-localized minimum and the 1$\sigma$ uncertainty contour. Our final result, 
$E_{\gamma}$ = 2779.0$^{+1.0}_{-0.8}$ keV and $\tau$ = 140$^{+50}_{-40}$ fs, is in agreement with the literature value, within uncertainties. Figures \ref{Fig3}(a)-(c) display similar analysis for the long-lived ($\tau>$3.5 ps) 2371-keV line. Here, the $\chi^2$ map shows a valley at $E_{\gamma}$ = 2370.6$^{+0.5}_{-0.3}$ keV, extending from 400 fs towards higher values, without a localized minimum. Both $^{19}$O results validate our novel Monte Carlo analysis.

We now turn to the lifetime of the 2$^+_2$ states in $^{20}$O and $^{16}$C. Figures \ref{Fig3} (g)-(i) refer to the 2$^+_2 \rightarrow 2^+_1$  decay in $^{20}$O. A well-defined minimum is found in the $\chi^2$ surface, yielding the values $E_{\gamma}$ = 2394.6$^{+1.0}_{-1.0}$ keV and $\tau$ = 150$^{+80}_{-30}$ fs. We note that the $\gamma$-ray energy agrees with the most precise value reported in literature, $E_{\gamma}$ = 2396(1) keV \cite{Wie05}, while the present determination of the lifetime is the first obtained thus far. For the $2^+_2\rightarrow 2^+_1$ decay, considering its 79(5)$\%$ branching ratio, here determined, one gets a partial lifetime of 190$^{+102}_{-40}$ fs. We stress that the above results rely on the AGATA excellent position resolution of the first $\gamma$ interaction point, determined with the combined use of Pulse Shape Analysis \cite{Ven04,Bru16} and the Orsay Forward Tracking algorithm \cite{Lop04}. Defining the $\gamma$ direction from the front segment centers only, as in the case of conventional Ge arrays, gives $\chi^2$ minima with much larger uncertainty (dashed lines in the insets of Fig. \ref{Fig3} (d)-(g)), making meaningless a comparison with theory. 

Figures \ref{Fig3} (j)-(k) report the analysis of the 2$^+_2$  state in $^{16}$C, performed on the $\gamma$-ray spectrum integrated over the entire angular range, i.e., {\ang{120}-\ang{180}}, due to the limited statistics. As a consequence, the 
$^{16}$C $\chi^2$ map, shown in Fig. \ref{Fig3}(j), displays a wide valley, resulting in a very limited sensitivity when $E_{\gamma}<$ 2216 keV. For 2$^+_2\rightarrow$2$^+_1$ transition energies $E_{\gamma}>$ 2216 keV, the procedure provides a more definite value, which would indicate a lifetime $\tau <$ 180 fs. Considering the most precise energy measurement $E_{\gamma}$ = 2217(2) keV \cite{Pet12}, there is a 78$\%$ probability for this scenario.

\begin{figure}[ht]
%\vspace*{0.1cm}
\resizebox{0.48\textwidth}{!}{\includegraphics{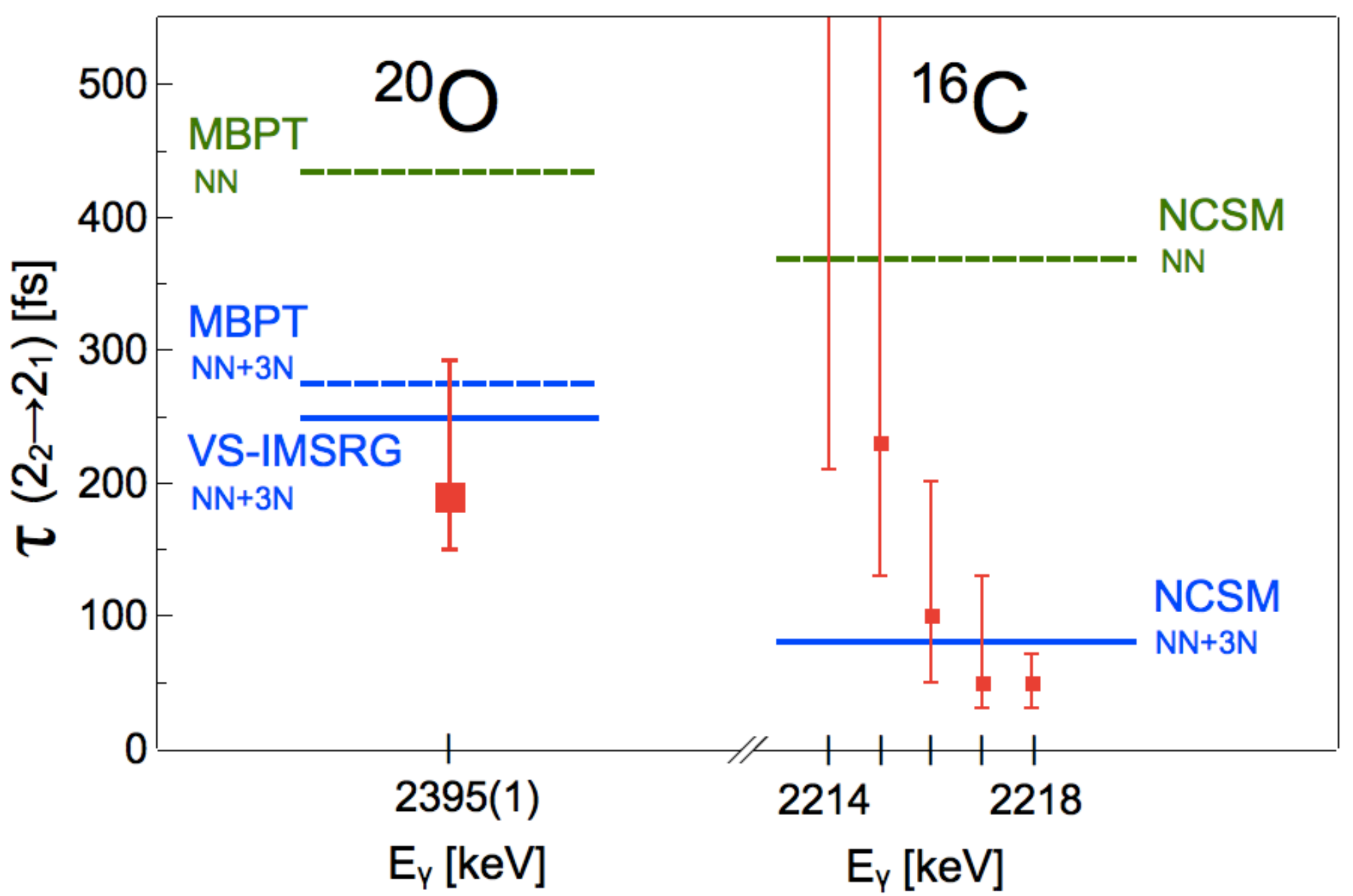}}
\caption{Partial lifetime for 2$^+_2\rightarrow$2$^+_1$ decays. \textit{Left:} $^{20}$O, experiment (symbol) compared to MBPT (dashed lines, with and without 3N interactions), and \textit{ab initio} VS-IMSRG (solid line) results. \textit{Right:} $^{16}$C, experiment (symbols)  for assumed $E_{\gamma}$ energies (see Fig. \ref{Fig3} j) and k)), including the uncertainty from a $\leq$13$\%$ branching ratio. Solid (dashed) lines show \textit{ab initio} NCSM predictions with (without) 3N interactions. The MBPT results use neutron effective charge e$_n$=0.4 and g-factors g$_s$=-3.55 and g$_l$=-0.09.} 
\label{Fig4}
\end{figure}

We performed calculations for $^{20}$O  by employing chiral NN interactions based on Ref. \cite{Ent03}, and 3N interactions fitted on top, considering only few-body systems up to $^{4}$He. 
First, the many-body perturbation theory (MBPT) valence-shell interactions from Ref. \cite{Heb15} were employed with an $^{16}$O core. NN and normal-ordered 3N interactions are included up to third order, neglecting residual 3N interactions. Effective operators are used to calculate EM transitions \cite{Gar15,Klo19}. The shell model diagonalizations were performed with the code ANTOINE \cite{Cau05}. 
The MBPT results reproduce well the 2$^+_1\rightarrow 0^+$ lifetime in $^{20}$O ($\tau$ = 11.7 ps vs. the experimental 
$\tau$ = 10.5(4) ps \cite{Pri16}), and this agreement does not depend significantly on the inclusion of 3N interactions. In turn, as shown in Fig. \ref{Fig4}, the calculated partial lifetime of the 2$^+_2\rightarrow 2^+_1$ decay, $\tau$ = 275 fs (dashed blue line), agrees well with the present measurement only when 3N interactions are considered (B(M1)=0.015 $\mu_N^2$, B(E2)=0.051 e$^2$fm$^4$, taking the experimental transition energy). When they are neglected, the 2$^+_2\rightarrow2^+_1$ partial lifetime is about 60$\%$ longer (dashed green line) and the energy of the 2$^+_2$ state is more than 1 MeV lower (see also Ref. \cite{Heb15}). This change is driven by the dominant (d$_{5/2}$)$^3$(s$_{1/2}$)$^1$ configuration, which is mostly missing in the NN calculation. 

Second, we performed \textit{ab initio} calculations with the VS-IMSRG \cite{Heb15, Tsu12, Str16, Str17, Str19}, based on the NN+3N interaction labeled EM1.8/2.0 in Ref. \cite{Sim17}. 
%A valence-shell interaction is decoupled for $^{20}$O in the sd shell, providing the energy of the $^{16}$O core so that total energies are obtained for all levels. 
The treatment of 3N interactions improves that of the MBPT, by including approximately the interactions between valence nucleons \cite{Str17}. In addition, we decouple valence-space operators consistent with the Hamiltonian as in Refs. \cite{Par17,Mor15}, avoiding the use of effective charges or g-factors and producing effective two-body EM operators. In the IMSRG(2) approximation used here, all operators are truncated at the two-body level, which leads to reduced B(E2) values as discussed in Ref. \cite{Par17}.
VS-IMSRG transition energies are in very good agreement with experiment, giving 1629 keV and 2422 keV for the 
2$^+_1\rightarrow$0$^+$ and 2$^+_2\rightarrow$2$^+_1$ decays, respectively. This is at variance with the MBPT results, which overestimate the experimental energies by about 400 keV. The VS-IMSRG 2$^+_2\rightarrow2^+_1$ partial lifetime in $^{20}$O, 
$\tau$ = 249 fs (solid line), also agrees very well with the present measurement (see Fig. \ref{Fig4}). B(M1)=0.0166 $\mu_N^2$ dominates over B(E2)=0.0684 e$^2$fm$^4$. The good agreement with the experimental lifetime also suggests a small impact of meson-exchange currents, not included in this calculation.

In $^{10-20}$C isotopes, Forss\'en et al. have performed NCSM calculations with NN+3N interactions \cite{For13}. Like in the VS-IMSRG, total energies and transition probabilities are obtained without effective charges or additional  parameters. Consistently with the MBPT calculation on $^{20}$O, Forss\'en et al. find the decay rates  of the 2$^+_2$, 3$^+_1$ and 4$^+_1$ excited states in $^{16}$C sensitive to 3N forces, with transition strengths reduced up to a factor 7. For the 
2$^+_2\rightarrow$2$^+_1$ decay, the NCSM finds that the B(M1)=0.063 $\mu_N^2$ dominates, yielding the partial lifetime 
$\tau$ = 81 fs. Figure \ref{Fig4} shows that the NCSM calculations are consistent with the experimental estimates for transition energy $E_{\gamma}\ge$ 2216 keV $-$ this scenario has 78$\%$ probability, as discussed above. The absence of a 3980-keV 2$^+_2\rightarrow$0$^+$ decay branch in our data, for which we extract an upper limit of 13$\%$ (compatible with the previous value of $\le$8.8$\%$\cite{Pet12}) is also consistent with the NCSM NN+3N results.  

Summarizing, we have measured, for the first time, the lifetime of the 2$^+_2$ state in $^{20}$O, and an estimate is given for the 2$^+_2$ state in $^{16}$C. To obtain such results, a novel Monte Carlo technique was developed to determine lifetimes in the tens-to-hundreds femtoseconds range, using low-energy transfer and deep-inelastic reactions. This technique will be essential for similar investigations in exotic neutron-rich nuclei produced with intense ISOL-type beams. Crucial in our work was the high precision provided by the AGATA $\gamma$-tracking array. The achieved results on transition probabilities agree well with predictions from MBPT and \textit{ab initio} VS-IMSRG for $^{20}$O, and NCSM calculations for $^{16}$C, showing that 3N interactions are needed to accurately describe electromagnetic observables in neutron-rich nuclei. From a broader perspective, the present measurement demontrates that high-precision $\gamma$ spectroscopy can benchmark
first principles nuclear structure calculations. The work paves the way toward comprehensive tests of \textit{ab initio} approaches, exploiting electromagnetic transitions in addition to standard comparisons based mostly on nuclear energies.

%\begin{center}
%\begin{strip}
%\begin{equation}
%H = 
%\left( \begin{array}{cccccc} 
%\varepsilon_{j} & 0 & \ldots & h(j,j'_1n_1J_1) & h(j,j'_2n_2J_2) & \ldots \\
%0 & \varepsilon_{j'''} & \ldots & h(j''',j'_1,n_1J_1) & h(j''',j'_2n_2J_2) & \ldots \\
%\ldots & \ldots & \ldots & \ldots & \ldots & \ldots \\
%h(j,j'_1n_1J_1)  & h(j''',j'_1,n_1J_1) & \ldots & \varepsilon_{j'_1} + \hbar\omega_{n_1J_1}
%& 0 & \ldots \\
%h(j,j'_2n_2J_2)  & h(j''',j'_2,n_2J_2) & \ldots & 0 & 
%\varepsilon_{j'_2} + \hbar\omega_{n_2J_2} & \ldots \\
%\ldots & \ldots & \ldots & \ldots & \ldots & \ldots \\
%\end{array} \right)
%\end{equation}
%%\end{strip}
%\end{center}

%\end{savenotes}

\bigskip

\begin{acknowledgements} We thank S. R. Stroberg for very useful discussions. This work was supported by the Italian Istituto Nazionale di Fisica Nucleare, by the Polish National Science Centre under Contract No. 2014/14/M/ST2/00738, 2013/08/M/ST2/00257 and 2016/22/M/ST2/00269, and by RSF Grant No. 19-42-02014. This project has received funding from the Turkish Scientific and Research Council (Project No. 115F103) and from the European Union’s Horizon 2020 research and innovation program under grant agreement No. 654002. This work was also supported by the Spanish Ministerio de Econom\'ia y Competitividad under
contract No. FPA2017-84756-C4-2-P and by the JSPS KAKENHI Grant No.18K03639, MEXT as “Priority issue on post-K computer”
(Elucidation of the fundamental laws and evolution of the universe), JICFuS, NSERC, the BMBF under Contract No.~05P18RDFN1 and the Deutsche Forschungsgemeinschaft (DFG, German Research Foundation) -- Projektnummer 279384907 -- SFB 1245 and the Cluster of Excellence PRISMA$^+$ EXC 2118/1 funded by DFG within the German Excellence Strategy (Project ID 39083149).
TRIUMF receives funding via a contribution through the National Research Council Canada.
\end{acknowledgements}

\end{document}